\documentclass[twocolumn]{aastex631}
\usepackage{booktabs}
\usepackage{amsmath}
\usepackage{graphicx}
\usepackage{subfigure}
\usepackage{longtable}

\shorttitle{}
\shortauthors{Taamoli et al.}

\def\photoz{photo-\textit{z}}

\begin{document}
\title{COSMOS2020: Disentangling the Role of Mass and Environment in Star Formation Activity of Galaxies at $0.4<z<4$}

\author[0000-0003-0749-4667]{Sina Taamoli}
\affiliation{Department of Physics and Astronomy, University of California, Riverside, 900 University Ave, Riverside, CA 92521, USA}

\author[0000-0002-7755-8649]{Negin Nezhad}
% \altaffiliation{ \texttt{Negin Khosravaninezhad and Sina Taamoli are equal contributors to this work and designated as co-first authors.}}
\affiliation{Department of Physics and Astronomy, University of California, Riverside, 900 University Ave, Riverside, CA 92521, USA}
\affiliation{Cosmic Dawn Center (DAWN), Denmark}
\affiliation{Niels Bohr Institute, University of Copenhagen, Jagtvej 128, DK-2200 Copenhagen, Denmark}

\author[0000-0001-5846-4404]{Bahram Mobasher}
\affiliation{Department of Physics and Astronomy, University of California, Riverside, 900 University Ave, Riverside, CA 92521, USA}

\author[0009-0008-4976-3216]{Faezeh Manesh}
\affiliation{Department of Physics and Astronomy, University of California, Riverside, 900 University Ave, Riverside, CA 92521, USA}

\author[0000-0003-3691-937X]{Nima Chartab}
\affiliation{Infrared Processing and Analysis Center, California Institute of Technology, Pasadena, CA 91125, USA}

\author[0000-0003-1614-196X]{John R. Weaver}
\affiliation{Department of Astronomy, University of Massachusetts, Amherst, MA 01003, USA}

\author[0000-0003-3578-6843]{Peter L. Capak}
\affiliation{Infrared Processing and Analysis Center, California Institute of Technology, Pasadena, CA 91125, USA}

\author[0000-0002-0930-6466]{Caitlin M. Casey}
\affiliation{Department of Astronomy, The University of Texas at Austin, 2515 Speedway Boulevard Stop C1400, Austin, TX 78712, USA}

\author[0000-0002-0236-919X]{Ghassem Gozaliasl}
\affiliation{Department of Computer Science, Aalto University, PO Box 15400, Espoo, FI-00 076, Finland}
\affiliation{Department of Physics, University of Helsinki, PO Box 64, FI-00014 Helsinki, Finland}

\author[0000-0002-9389-7413]{Kasper E. Heintz}
\affiliation{Cosmic Dawn Center (DAWN), Denmark}
\affiliation{Niels Bohr Institute, University of Copenhagen, Jagtvej 128, DK-2200 Copenhagen, Denmark}
\affiliation{Department of Astronomy, University of Geneva, Chemin Pegasi 51, 1290 Versoix,
Switzerland.}

\author[0000-0002-7303-4397]{Olivier Ilbert}
\affiliation{Aix Marseille Univ, CNRS, LAM, Laboratoire d’Astrophysique de Marseille, Marseille, France}

\author[0000-0001-9187-3605]{Jeyhan S. Kartaltepe}
\affiliation{Laboratory for Multiwavelength Astrophysics, School of Physics and Astronomy, Rochester Institute of Technology, 84 Lomb Memorial Drive, Rochester, NY
14623, USA}

\author[0000-0002-9489-7765]{Henry J. McCracken}
\affiliation{Institut d’Astrophysique de Paris, UMR 7095, CNRS, and Sorbonne Université, 98 bis boulevard Arago, F-75014 Paris, France}

\author[0000-0002-1233-9998]{David B. Sanders}
\affiliation{Institute for Astronomy, University of Hawaii, 2680 Woodlawn Drive, Honolulu, HI 96822, USA}

\author[0000-0002-0438-3323]{Nicholas Scoville}
\affiliation{California Institute of Technology, 1200 E. California Boulevard, Pasadena, CA 91125, USA}

\author[0000-0003-3631-7176]{Sune Toft}
\affiliation{Cosmic Dawn Center (DAWN), Denmark}
\affiliation{Niels Bohr Institute, University of Copenhagen, Jagtvej 128, DK-2200 Copenhagen, Denmark}

\author[0000-0002-4465-8264]{Darach Watson}
\affiliation{Cosmic Dawn Center (DAWN), Denmark}
\affiliation{Niels Bohr Institute, University of Copenhagen, Jagtvej 128, DK-2200 Copenhagen, Denmark}

\begin{abstract}
The role of internal and environmental factors in the star formation activity of galaxies is still a matter of debate, particularly at higher redshifts. Leveraging the most recent release of the COSMOS catalog, COSMOS2020, and density measurements from our previous study we disentangle the impact of environment and stellar mass on the star formation rate (SFR), and specific SFR (sSFR) of a sample of $\sim 210,000$ galaxies within redshift range $0.4< z < 4$ and present our findings in three cosmic epochs: 1) out to $z\sim 1$, the average SFR and sSFR decline at extremely dense environments and high mass end of the distribution which is mostly due to the presence of the massive quiescent population; 2) at $1<z<2$, the environmental dependence diminishes, while mass is still the dominant factor in star formation activity; 3) beyond $z\sim 2$, our sample is dominated by star-forming galaxies and we observe a reversal of the trends seen in the local universe: the average SFR increases with increasing environmental density. Our analysis shows that both environmental and mass quenching efficiencies increase with stellar mass at all redshifts, with mass being the dominant quenching factor in massive galaxies compared to environmental effects. At $2<z<4$, negative values of environmental quenching efficiency suggest that the fraction of star-forming galaxies in dense environments exceeds that in less dense regions, likely due to the greater availability of cold gas, higher merger rates, and tidal effects that trigger star formation activity.
\end{abstract}


\section{Introduction}\label{sec:introduction}

The interplay between galaxies' environments and internal processes shapes their evolution, with these factors being interdependent. Over the past decades, the study of these influences across different cosmic epochs and mass regimes has significantly advanced through the advent of multi-wavelength, deep, and wide-area galaxy surveys spanning a broad range of redshifts.

Environmental factors are shown to influence star formation activity \citep{Scoville2013,darvish2016,chartab2020,taamoli2024}, morphology \citep{Mandelbaum2006,capak2007,Bamford2009}, metal enrichment \citep{sattari2021,chartab2021B}, and quenching mechanisms \citep{peng2010,Poggianti2017,zheng2024}. 

In the local universe, denser environments (e.g., galaxy clusters and filaments) are predominantly populated with early-type, red, passive galaxies, whereas less dense environments mainly host late-type, blue, star-forming galaxies \citep{dressler1980, balogh2004, kauffmann2004, baldry2006, Bamford2009, peng2010, woo2013}. In other words, both star formation rate (SFR) and specific star formation rate (sSFR) are relatively lower in high-density environments compared to their counterparts in lower densities, known as ``field galaxies'' \citep{darvish2014,chartab2020,taamoli2024}. This is partly attributed to environmental quenching mechanisms, which may involve one or a combination of processes such as ram pressure stripping \citep{gunn_infall_1972,Moore1999, Brown2017,Barsanti2018}, strangulation or starvation \citep{Moore1999,peng2015}, galaxy harassment \citep{moore_galaxy_1996,farouki_computer_1981}, galaxy-galaxy interactions \citep{Tanaka2004}.

Internal processes also critically affect star formation activity and are key drivers of galaxy quenching (mass quenching) \citep{peng2010,Schawinski2014}. While environmental quenching mostly depends on the local density of galaxies, mass quenching primarily depends on the stellar mass ($M_{\star}$) and the morphological type of galaxies \citep{peng2010}. For low-mass galaxies ($\log(M_{\star} / M_{\odot}) < 10$), gas outflows caused by stellar feedback mechanisms, such as stellar winds or supernova explosions, significantly impact the suppression of star formation \citep{Larson1974, dekel1986, dallavecchia2008}. For galaxies with higher stellar masses ($\log(M_{\star} / M_{\odot}) \gtrsim 10$), feedback from active galactic nuclei (AGN), morphological quenching \citep{Martig2009}, and shock heating are likely to play a more significant role in the cessation of star formation \citep{Croton2006,Nandra_2007,fabian2012,Cicone,Bremer2018}. We should note that mass build-up of galaxies depends on the environment, and the galaxy's mass, in turn, determines its gravitational potential well, which regulates interactions with the environment, including infall/outflow rates and gas accretion. Low-mass galaxies have shallower potential wells and are more susceptible to processes such as ram pressure stripping, while high-mass systems, with deeper potential wells and higher binding energies in their interstellar medium (ISM) and circumgalactic medium (CGM), are more affected by mechanisms such as shock heating \citep{Dekel2006,Goubert2024}. 

Beyond $z\sim 1$, the relation between SFR/sSFR and environmental density remains a topic of ongoing investigation, with studies reporting varying or even contradictory results. While some studies suggest that the trends at higher redshifts align with those observed in the local universe \citep{patel2009,Muzzin2012,chartab2020}, others have found no significant correlation or weakening of the local trends at $z>1$  and argue that most of the observed trends at higher redshifts are due to the cosmic variance, lack of extremely dense environments in COSMOS field (small dynamical range), or selection effects \citep{Grutzbauch2011,Scoville2013, darvish2016}. Conversely, several studies have observed a reversal of the local universe trends at $z>1$, indicating that galaxies in denser environments were, on average, more star-forming than those in less dense regions \citep{Ideue2012, Welikala2016,lemaux_vimos_2022, taamoli2024}. This reversal could be due to the ample gas supply in dense environments that fuels star formation, along with higher merger rates that could trigger star formation in these settings. 

Given the inconsistencies among high-redshift results, further analysis of wide and deep galaxy surveys spanning a broad range of environments is essential to disentangle the contributions of mass and environment to star formation activity. The COSMOS2020 catalog, with its multi-band photometric data across 1.7 deg\(^2\), offers a valuable opportunity to address this issue \citep{weaver2022}. Although the COSMOS field lacks extremely high overdensities up to \(z \sim 2\), it includes a variety of structures, enabling the study of galaxies in diverse environments.

Building on a recent study where we reconstructed the environmental density field of galaxies across \(0.4<z<6\) using COSMOS2020 data \citep{taamoli2024}, this work investigates the relative contributions of mass and environment to galaxy quenching and star formation activity and the influence of quiescent galaxies on the observed trends between mass, star formation activity, and environment.

The paper is organized as follows: In Section \ref{sec:data} we briefly review the measurement of the large-scale structures at different redshifts in the COSMOS2020 field and our selection criteria. In Section \ref{sec:results}, we present our analysis of the correlation between mass, environment, and star formation activity as a function of redshift. In Section \ref{sec:comparison}, we discuss the results and compare them with previous studies, and in Section \ref{sec:summary}, we summarize our findings.

Throughout this work, we assume a flat $\Lambda\text{CDM}$ cosmology with $H_{0} = 70 \, \text{km} \, s^{-1} \text{Mpc}^{-1}$ , $\Omega_{m0} = 0.3$, and $\Omega_{\Lambda0}= 0.7 $. All magnitudes are expressed in the AB system and the physical parameters are measured assuming a Chabrier initial mass function.

\section{Data} \label{sec:data}

We utilize a combination of photometric measurements from \texttt{The Farmer} (\citealt{weaver_farmer_2023}a, \citealt{farmer2023}b) and physical parameters derived using the \texttt{Lephare}\footnote{\href{https://www.cfht.hawaii.edu/~arnouts/LEPHARE/lephare.html}{https://www.cfht.hawaii.edu/\(\sim\)arnouts/LEPHARE/lephare.html}} SED fitting code \citep{ilbert2006, Arnouts2002lephare}, both sourced from the COSMOS2020 catalog \citep{weaver2022}. For environmental densities, we rely on the catalog from \citep{taamoli2024}, which uses photometric redshifts (hereafter \photoz) and their probability distributions (zPDF) from the \texttt{Farmer+Lephare} combination to produce a density catalog for approximately 210,000 galaxies at \(0.4 < z < 5\) using adaptive Kernel Density Estimation (wKDE) method. Following the procedure in \citep{ilbert2013}, galaxies are classified as star-forming or quiescent based on their rest-frame \texttt{NUV-r} and \texttt{r-J} colors. The selection criteria for our sample are as follows:

\begin{itemize}
    \item Estimated \photoz, ``\texttt{lp\_zPDF}'', in the range of $0.4<z<6$.
    \item Sources with large uncertainties in their photo-\textit{z} measurements ($\Delta z > 2$) are filtered out. $\Delta z$ is the 68\% confidence interval on estimated \photoz.
    \item $1.604<\delta<2.817$ and $149.398<\alpha<150.787$: This is the largest region containing spatially homogeneous selection function and depth.
    \item $K_{S}=24.5$ (AB) magnitude cut on Ultra-Vista near-infrared data.
\end{itemize}

The rationale behind these selection criteria is explained in detail in \cite{taamoli2024}. For stellar masses and (specific) star formation rates used in the following section, we use median values of likelihood functions for these parameters reported in COSMOS2020 (``\texttt{lp\_mass\_med}'', ``\texttt{lp\_SFR\_med}'', and ``\texttt{lp\_sSFR\_med}'').

\section{Results} \label{sec:results}
In this section, we examine the relationships between stellar mass (hereafter SM), SFR, and environment in galaxies (Section \ref{subsec:sfr_m_env}), followed by an analysis of environmental and mass quenching efficiency as a function of SM and its evolution over look-back time (Section \ref{subsec:quenching}).

\subsection{star formation activity vs. Environment} \label{subsec:sfr_m_env}

\begin{figure*}[htp]
    \centering
    \begin{subfigure}
        \centering
        \includegraphics[width=0.98\textwidth]{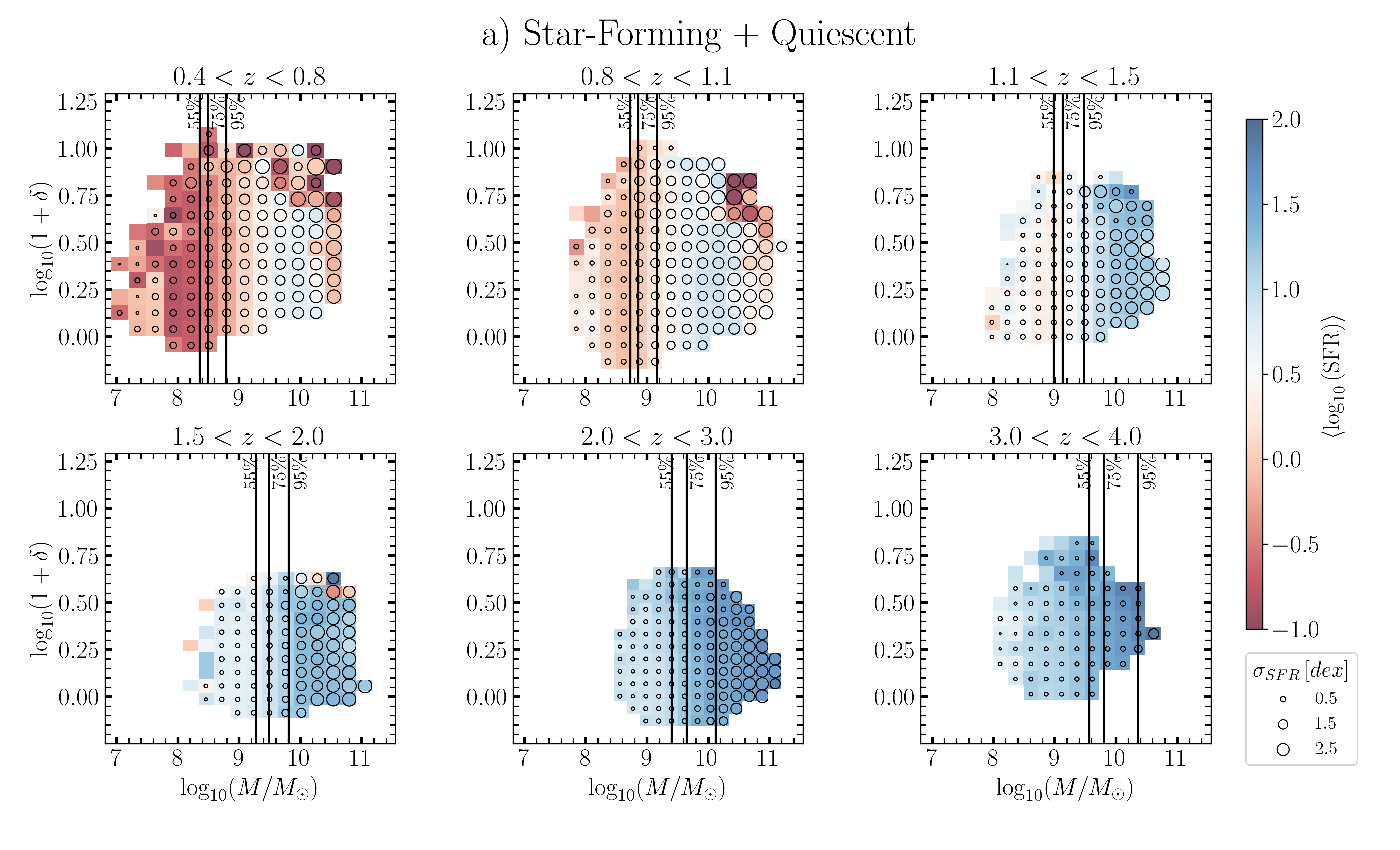}
    \end{subfigure} 
    \begin{subfigure}
        \centering
        \includegraphics[width=0.98\textwidth]{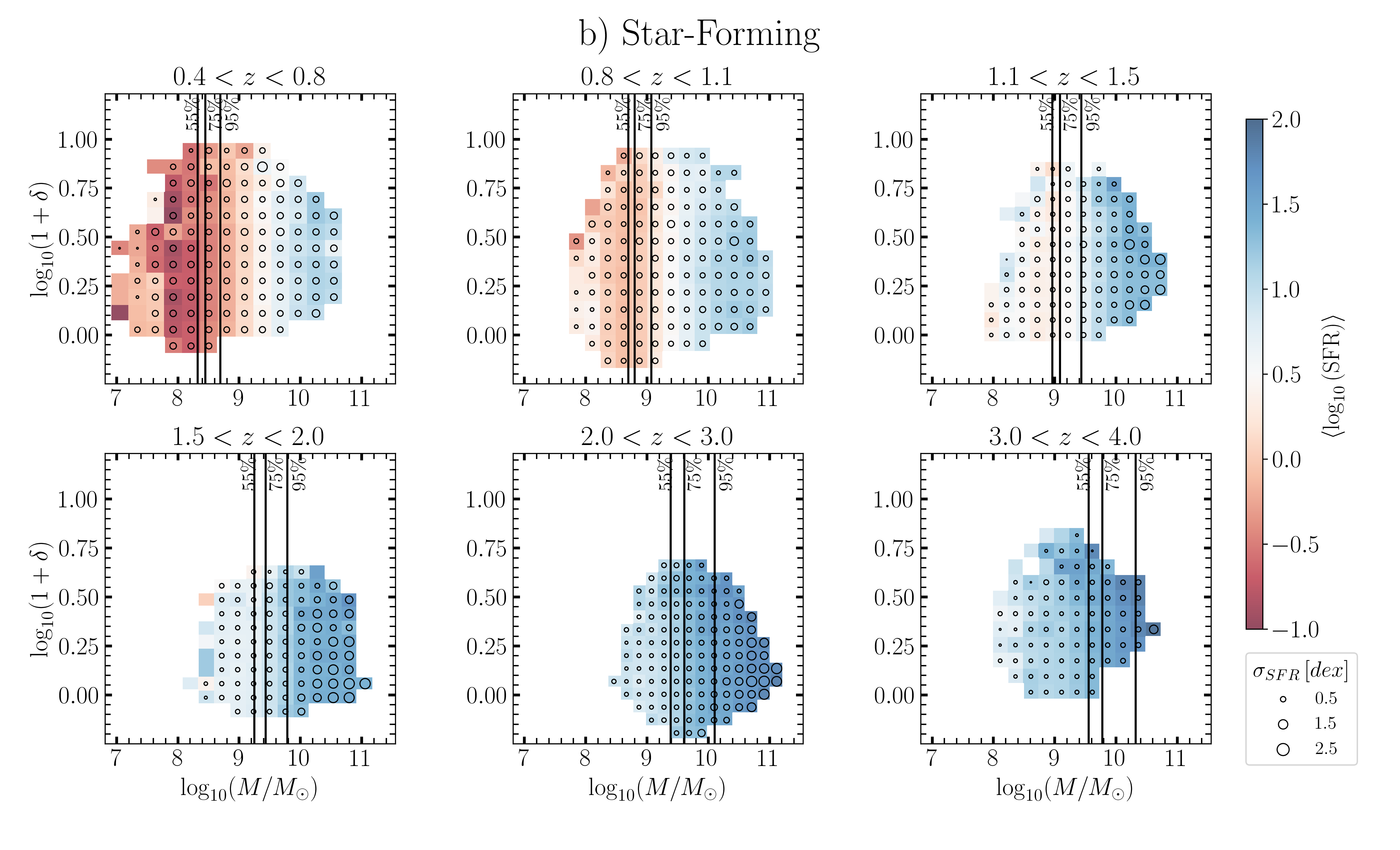}
    \end{subfigure}
    \caption{Median of $\log$(SFR) in bins of SM and environmental density, $\log(1+\delta)$, in the overall sample of galaxies (panel a), and for star-forming galaxies (panel b) in 6 redshift intervals spanning $0.4<z<4$. Vertical lines show 55\%, 75\%, and 95\% mass completeness limit. Circles show the standard deviation of SFR in each bin.}
    \label{fig:d-m-sfr}
\end{figure*}

\begin{figure*}[t!]
    \centering
    \begin{subfigure}
        \centering
        \includegraphics[width = 0.98\textwidth]{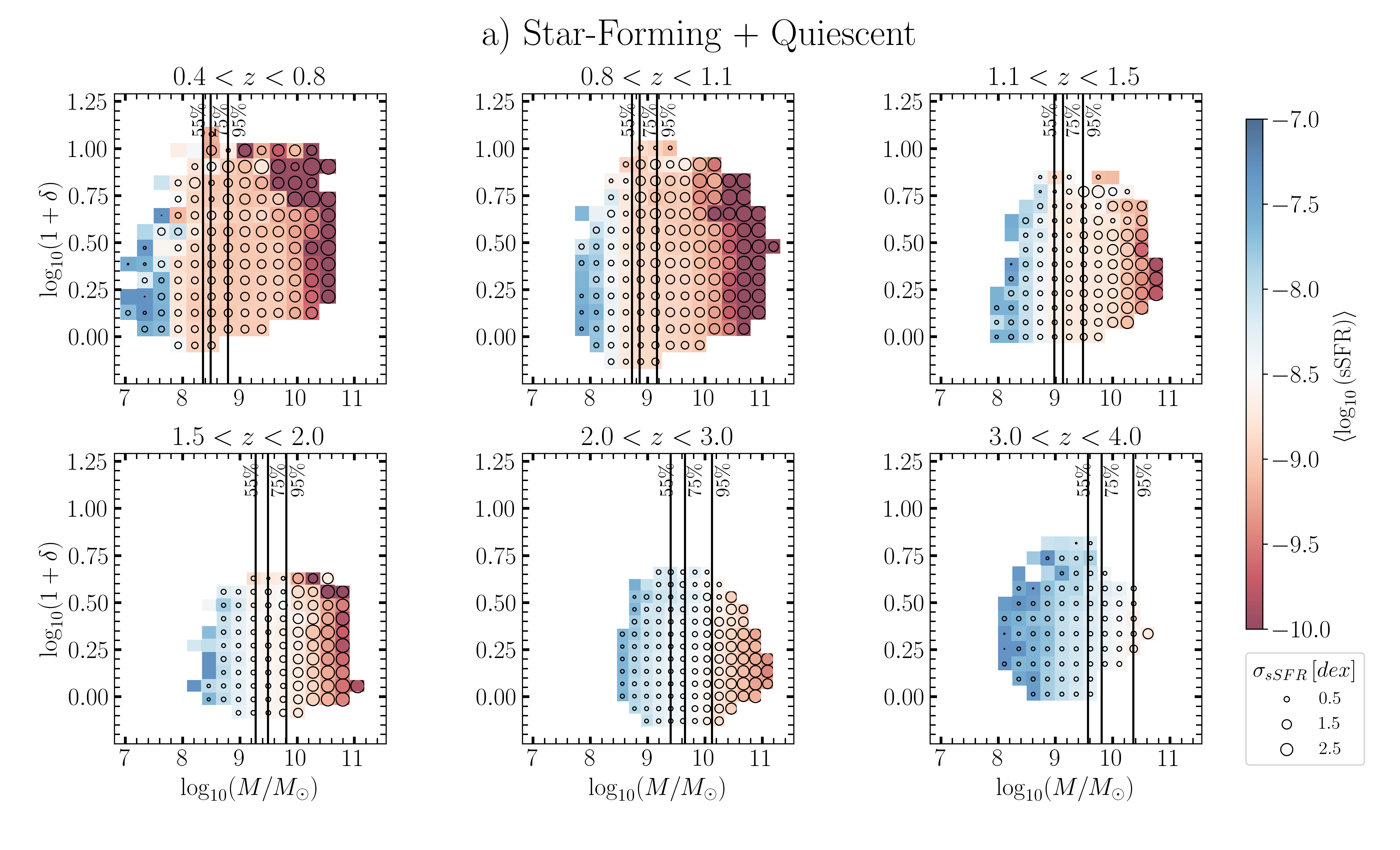}
    \end{subfigure}
    \begin{subfigure}
        \centering
        \includegraphics[width = 0.98\textwidth]{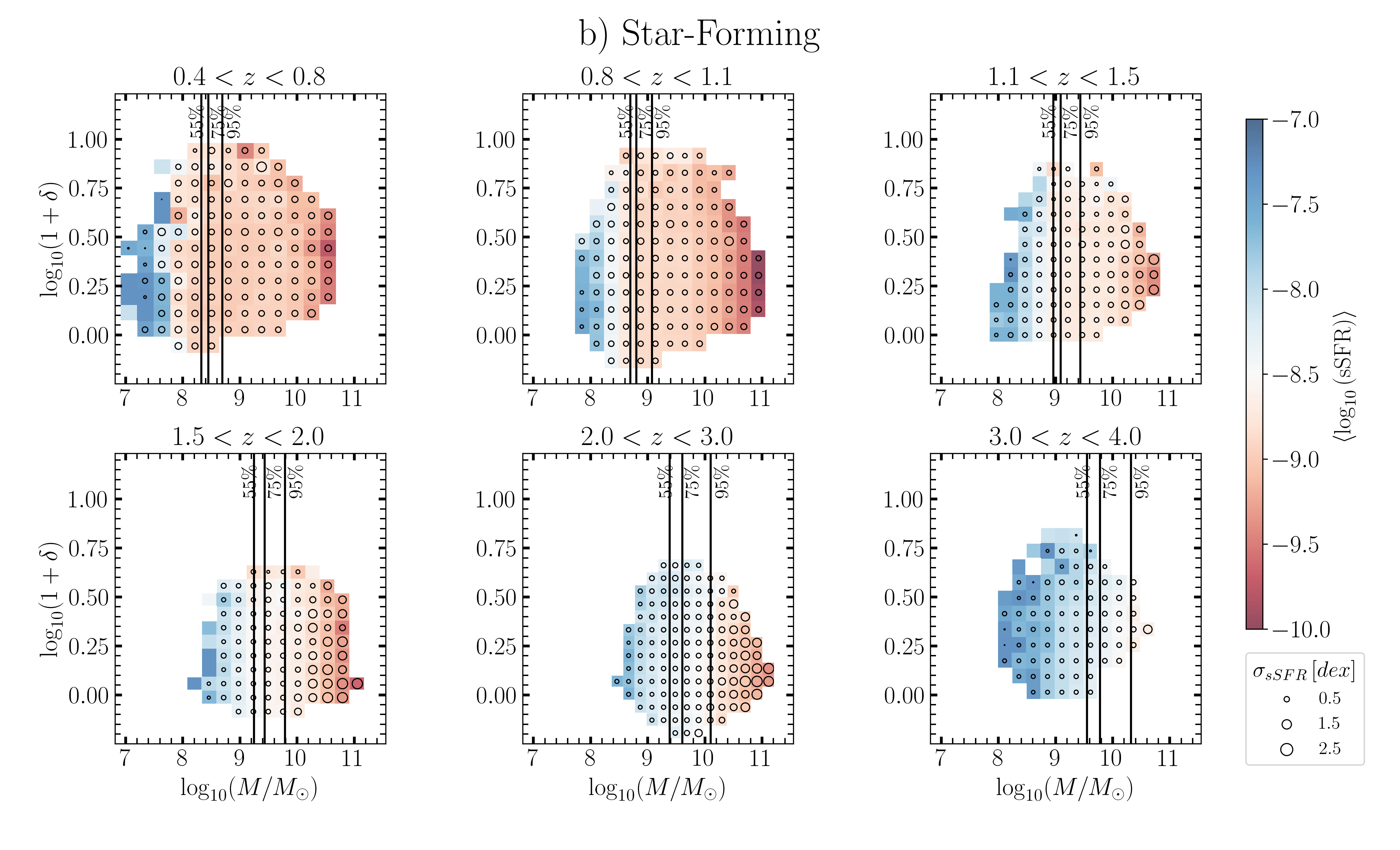}
    \end{subfigure}
        \caption{Median of $\log$(sSFR) in bins of stellar mass and environmental density, $\log(1+\delta)$, in the overall sample of galaxies (panel a), and for star-forming galaxies (panel b) in 6 redshift intervals spanning $0.4<z<4$. Vertical lines show 55\%, 75\%, and 95\% mass completeness limit. Circles show the standard deviation of SFR in each bin.}
    \label{fig:d-m-ssfr}
\end{figure*}

Figure \ref{fig:d-m-sfr} presents the median SFR in bins of SM and environmental density, $\log(1+\delta)$, for two samples: 1) a sample of star-forming and quiescent galaxies (overall sample), shown in panel a; 2) a sample of only star-forming galaxies, shown in panel b. Bins containing fewer than 20 galaxies are shown in white in all figures. Vertical lines represent 55\%, 75\%, and 95\% mass completeness limits, respectively from left to right. Completeness limits are calculated following the same procedure outlined in (\citealt{peng2010}; also summarized in \citealt{taamoli2024}). To better justify the significance of the binned statistics, we show the scatter of SFR in each bin with circles showing the standard deviation in that bin: bigger circles indicate a larger scatter around the median value. Figure \ref{fig:d-m-ssfr} supplements these observations by depicting the median sSFR in bins of mass and environmental density. 

To quantify the findings presented in this section, we report Spearman partial correlation coefficients for the star-forming, quiescent, and overall samples, analyzing the relationships between SFR/sSFR and the environment, conditioned by SM, in each redshift interval. This statistical method extends Spearman's rank correlation by isolating the direct relationship between two variables while accounting for the influence of other confounding factors \citep{Spearman1904}. 
The $P$-value associated with each correlation coefficient indicates the probability of observing the result under the null hypothesis of no correlation. A $P$-value below 5\% ($P<0.05$) signifies statistical significance, while values greater than 5\% suggest the correlation may be due to random chance. Correlation coefficients with $P$-values $>$ 5\% are marked with a star in Table \ref{table:corrcoeffs}. In the following paragraphs, we present our findings in three cosmic epochs, according to Figure \ref{fig:d-m-sfr}, Figure \ref{fig:d-m-ssfr}, and Table \ref{table:corrcoeffs}:

%%%%%%%%%%%%%%%%%%% LOW REDSHIFT %%%%%%%%%%%%%%%%%%%%%%%%%%%
At lower redshifts, out to $z\sim 1$, the overall sample shows that the SFR is, on average, smaller in high-density environments and for massive galaxies (Fig \ref{fig:d-m-sfr}, panel a). The environmental dependence of SFR is almost negligible for low-mass galaxies ($\log(M/M_{\odot}) \lesssim 10$), except for high-density environments ($\log(1+\delta) \gtrsim 0.75$) which host a larger fraction of quiescent galaxies (comparison between Fig \ref{fig:d-m-sfr}, panels a and b), resulting in lower median SFR. This indicates that environmental quenching mechanisms cease SFR in extremely dense environments. SFR is generally more sensitive to SM: at a given density, the average SFR increases with increasing SM, probably due to the availability of a larger gas reservoir fueling star formation activity and aiding the mass build-up in galaxies. There is a steep drop in the median SFR of the most massive galaxies in our sample, which disappears in Fig \ref{fig:d-m-sfr}, panel b, where quiescent galaxies are removed from the sample. Consequently, this sharp decline in SFR, along with the large standard deviation at the high-mass end of the overall sample, is attributed to the presence of the massive quiescent population. These massive galaxies ($\log(M/M_{\odot})\gtrsim 10.5$), have consumed most of their gas content to form stars and assemble their mass, resulting in relatively smaller SFRs. Similar trends are observed for sSFR in figure \ref{fig:d-m-ssfr}: there is a steep decrease in sSFR in massive galaxies ($\log(M/M_{\odot}) \gtrsim 10$), accompanied by large standard deviations in the overall sample (Fig. \ref{fig:d-m-ssfr}, panel a) which is mostly due to the presence of quiescent population. In addition to the mass dependence, an environmental trend is seen at $0.4<z<0.8$ for less massive galaxies ($\log(M/M_{\odot}) \lesssim 9.5-10$) in Figure \ref{fig:d-m-ssfr}, panel a: lower average sSFR and higher standard deviation at high-density environments ($\log(1+\delta) \gtrsim 0.75$). The star-forming sample (panel b), shows the same trend but less steep and with smaller standard deviations compared to the overall sample (panel a). Panel b of figure \ref{fig:d-m-ssfr} shows that even when we remove the quiescent population from the sample, sSFR drops by a factor of $\sim 1.5 \, dex$ at the high mass end of the distribution ($\log(M/M_{\odot})>10.5$). According to Fig. \ref{fig:d-m-ssfr}, panel b, there is a very weak environmental dependence in the star-forming sample within this redshift range. Correlation coefficients reported in Table \ref{table:corrcoeffs}, confirm our interpretation: In general, there is a negative correlation between environment and star formation activity up to $z \sim 1$ in all populations (overall, star-forming, and quiescent), but the environmental dependence of both SFR and sSFR is smaller in the star-forming sample compared to the overall sample, with this difference being more significant at $0.8<z<1.1$, compared to $0.4<z<0.8$.
%%%%%%%%%%%%%%%%%%% INTERMEDIATE REDSHIFT %%%%%%%%%%%%%%%%%%%
At intermediate redshifts ($1 < z < 2$), the environmental dependence of median SFR and sSFR is much shallower compared to the lower redshifts (Figures \ref{fig:d-m-sfr} and \ref{fig:d-m-ssfr}). At $1.1<z<1.5$, we report a mass dependence for SFR in both overall and star-forming samples, with the latter being steeper. In this range, the median SFR increases by $\sim 0.5 \, dex$ as mass increases from $\log(M/M_{\odot})\sim 9.5$ to $10.5$. Larger standard deviations in the overall sample (Fig.~\ref{fig:d-m-sfr}, panel a), which disappear in the star-forming sample (Fig.~\ref{fig:d-m-sfr}, panel b), are due to the presence of a quiescent population at the massive end, although these differences are smaller compared to lower redshifts. In contrast to lower redshifts, we do not observe a severe decline in average SFR at the high mass-end of the distribution in the overall sample (Fig.~\ref{fig:d-m-sfr}, panel a). This epoch is referred to as ``cosmic noon'' where the universal star formation activity reaches to its highest level. Unlike SFR, the median sSFR decreases by a factor of $\sim 1 \, dex$ in the overall sample (Fig.~\ref{fig:d-m-sfr}, panel a) and $\sim 0.5 \, dex$ in the star-forming sample by mass increasing from $\log(M/M_{\odot})\sim 9.5$ to $10.9$.  In general, no significant environmental dependence is observed for either SFR or sSFR at this redshift. Correlation coefficients, reported in Table~\ref{table:corrcoeffs}, confirm this interpretation: at $1.1<z<1.5$, in both overall and star-forming samples, correlation coefficients are smaller compared to the lower redshifts and the associated $P$-values are greater than $5\%$, indicating that the trends cannot be confidently explained by a monotonic relationship. At $1.5<z<2$, we find a negative correlation coefficient but still weaker than lower redshifts. In general, with smaller correlation coefficients and larger $P$-values at $1<z<2$, we must be cautious in our interpretations. We consider this redshift range a transitional epoch between the trends observed at lower ($z<1$) and higher redshifts ($z>2$).

\begin{figure*}
    \centering
    \includegraphics[width = 1\textwidth]{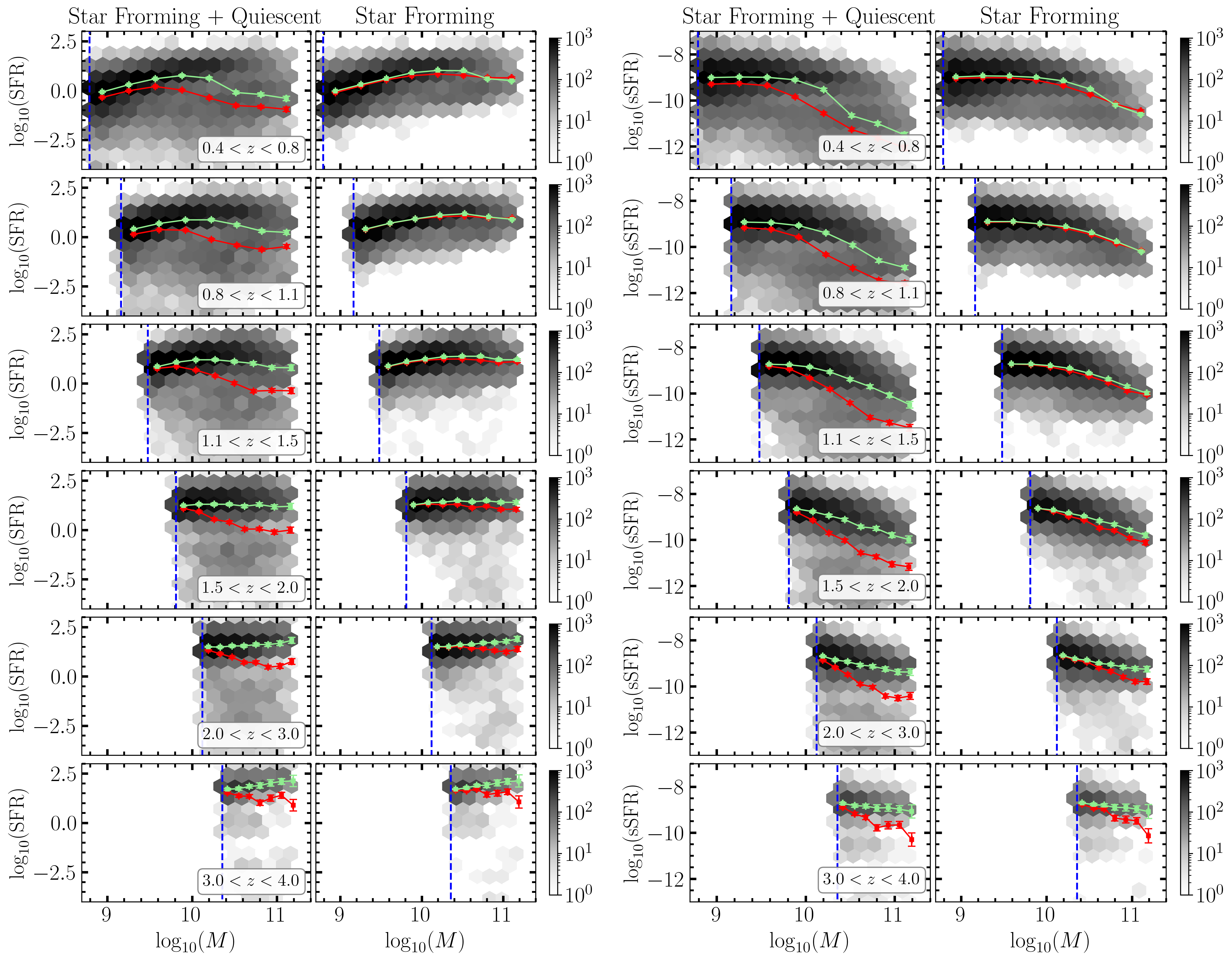}
    \caption{First two columns: SFR as a function of stellar mass for the overall (column 1) and star-forming (column 2) samples. The last two columns depict sSFR as a function of stellar mass for the overall (column 3) and star-forming (column 4) samples. The green and red curves represent the median and mean values of SFR and sSFR in bins of stellar mass, respectively. Error bars represent the standard deviation of the mean value. Vertical dashed lines show the 95\% completeness limit in each redshift interval.}
    \label{fig:binnedstat}
\end{figure*}

%%%%%%%%%%%%%%%%%%% HIGH REDSHIFT %%%%%%%%%%%%%%%%%%%%%%%%%%%
At higher redshifts ($2 < z < 4$), the average SFR is more uniform among galaxies with different masses and in various environments. In the highest redshift bin ($3< z< 4$), only a small fraction of galaxies are above the 95\% completeness limit, however, if we consider lower completeness limits (e.g., 75\%) there is a gentle increase in average SFR as we go to higher masses, in both overall and star-forming samples (Fig.~\ref{fig:d-m-sfr}, panels a and b). At this redshift range, the standard deviations are generally smaller than in other redshift ranges for both the overall and the star-forming sample (Fig.~\ref{fig:d-m-sfr}, panel a and b). According to \citep{taamoli2024}, the fraction of quiescent population fraction is $\sim 14$-$19\%$ in all redshift bins, and this fraction is at its lowest in the two highest redshift intervals ($\sim 14\%$), therefore the smaller difference between standard deviations in the overall and star-forming samples is due to the smaller population of quiescent galaxies and uniformity of SFR across both star-forming and quiescent population. Additionally, we are biased towards the most massive and luminous galaxies, which likely exhibit higher star formation activity. The same observation applies to the sSFR-mass-environment relation (Fig.~\ref{fig:d-m-ssfr}). At $2<z<3$, the median sSFR decreases by $\sim 0.5 \, dex$ as mass increases from $\log(M/M_{\odot})\sim 10$ to $11$. Furthermore, there is a gentle diagonal trend indicating that median sSFR decreases by decreasing environmental density. Correlation coefficients reported in Table \ref{table:corrcoeffs} show a positive correlation between star formation activity (SFR and sSFR) and environmental density in both the overall and star-forming samples, with smaller differences between these two samples compared to lower redshifts. However, these correlations are difficult to be visually spotted in Figures~\ref{fig:d-m-sfr} and \ref{fig:d-m-ssfr}.

In general, both panels a and b in figures \ref{fig:d-m-sfr} and \ref{fig:d-m-ssfr}, show that star formation activity increases with increasing redshift (see also \citealt{taamoli2024}; Figure 10). In addition, the standard deviation within bins of the overall sample (panel a) is generally larger than that in the star-forming sample (panel b), primarily due to the presence of quiescent galaxies, which contribute to the increased scatter around the average SFR/sSFR.

Figure~\ref{fig:binnedstat} shows SFR/sSFR (columns 1 \& 2 / columns 3 \& 4) as a function of stellar mass for both overall and star-forming samples (columns 1 \& 3 / columns 2 \& 4) across six redshift intervals. The gray-scaled map shows the population distribution across SFR-SM space. Hexbins with a population of fewer than five galaxies are set to zero. Green and red curves represent the median and mean values of the SFR/sSFR in bins of SM, with error bars representing the standard error of the mean. The mean values are shown to better illustrate the effect of the quiescent population on the observed trends. Vertical dashed lines denote the 95\% mass completeness limit. 

\begin{table*}[ht]
\centering
\begin{minipage}{0.83\linewidth}
\caption{Properties of subsamples across six redshift ranges, along with partial correlation coefficients for both the overall and star-forming samples. Correlation coefficients with $P$-values greater than 5\% are marked with $^{*}$ to indicate a lack of statistical significance.} 
\vspace{0.1in}
\end{minipage}
\centering
    \begin{tabular}{| l | l l | l l l | l l l |}
    \hline\hline
    Redshift Range & $\log(\text{M}_{min}/\text{M}_{\odot})$ & Size & \multicolumn{3}{c|}{SFR vs. density} & \multicolumn{3}{c|}{sSFR vs. density} \\
     & & & (SF+Q) & SF & Q & (SF+Q) & SF & Q \\
    [0.5ex]
    \hline
    % $***^{\star}$\footnotesize(P=***)
    $0.4<z<0.8$ & 8.791 & 26548 & $-0.064$ & $-0.023$ & $-0.088$ & $-0.050$ & $-0.019$ & $-0.072$\\ 
    $0.8<z<1.1$ & 9.163 & 23561 & $-0.051$ & $-0.006^{\star}$ & $-0.159$ & $-0.041$ & $-0.008^{\star}$ & $-0.150$\\
    $1.1<z<1.5$ & 9.479 & 22643 & $-0.002^{\star}$ & $0.009^{\star}$ & $-0.037$ & $0.004^{\star}$ & $0.009^{\star}$ & $-0.047$ \\
    $1.5<z<2$   & 9.811 & 13099 & $-0.032$ & $-0.007^{\star}$ & $-0.071$ & $-0.024$ & $-0.014$  & $-0.070$\\
    $2<z<3$     & 10.122& 9234  & $0.070$ & $0.072$ & $-0.024^{\star}$ & $0.073$ & $0.071$  & $-0.031^{\star}$\\
    $3<z<4$     & 10.357& 2162  & $0.125$ & $0.126$ & $0.040^{\star}$ & $0.134$ & $0.135$  & $0.074^{\star}$\\ [1ex]
    \hline
    \end{tabular}
\label{table:corrcoeffs}
\vspace{-0.13in}
\parbox{0.85\linewidth}{\flushleft $^{\star}$ $P>5\%$}
\end{table*}

By comparing columns 1 and 2 for SFR, and columns 3 and 4 for sSFR, we conclude that at all redshifts, the quiescent population influences the trends by lowering both SFR and sSFR, especially at higher masses ($\log(M/M_{\odot})>10$). Up to $z \sim 1.5-2$, quiescent galaxies are primarily responsible for lowering SFR and sSFR in the overall sample, particularly at higher masses. At higher redshifts ($z > 2$), there remains a population of low-SFR/sSFR galaxies even after excluding those flagged as quiescent. In the following paragraphs, we interpret the effect of quiescent population on the SFR (columns 1 and 2), and sSFR (columns 3 and 4), at different redshifts.

%%%%%%%%%%%%%% Fig 3: Low redshift %%%%%%%%%%%%%%%%%%
Out to $z \sim 1$, and in the star-forming sample (column 2), SFR increases by $\sim 0.5$-$0.75 \, dex$ as mass increases from $\log(M/M_{\odot})\gtrsim 9.5$ to $11$. Column 1 shows that the quiescent population lowers both mean and median SFR, particularly at ($\log(M/M_{\odot})\gtrsim 10.5$). As shown in columns 3 and 4, sSFR decreases significantly for masses higher than $\log(M/M_{\odot})\gtrsim 10.2$-$10.5$. Similar to our observation in the SFR-mass relation, the quiescent population steepens this trend at the massive end of the distribution ($\log(M/M_{\odot}) \gtrsim 10.5$). The differences between the overall and star-forming samples can be better seen in the mean values (red curves). 

%%%%%%%%%%%%%% Fig 3: mid redshift %%%%%%%%%%%%%%%%%%
At ($1 < z < 2$), the quiescent population does not significantly affect the median SFR and sSFR across the entire mass range. However, the mean values of SFR and sSFR are smaller in the overall sample as we go to the higher masses. In contrast to lower redshifts ($z \lesssim 1.1$), there remains a population of low-SFR galaxies that are flagged as quiescent and remain in the star-forming sample. This larger scatter causes the mean SFR and sSFR to deviate from the median value at higher masses. Both mean and median sSFR decrease monotonically with increasing mass.

%%%%%%%%%%%%%% Fig 3: high redshift %%%%%%%%%%%%%%%%%
At ($2 < z < 4$), SFR monotonically increases by $\sim 1.5 \, dex$ from $\log(M/M_{\odot}) \gtrsim 10.2$ to $10.9$. However, sSFR decreases monotonically by increasing mass. Similar to intermediate redshifts, a low-SFR population persists in the star-forming sample even after excluding the quiescent population from the overall sample.

\subsection{Mass and Environmental Quenching Efficiency} \label{subsec:quenching}

To assess the individual contributions of stellar mass and environmental factors to galaxy quenching, we report mass quenching efficiency and environmental quenching efficiency employing the definition provided by \citep{chartab2020} based on the framework outlined in \cite{peng2010}. Each metric aims to isolate the influence of one factor while holding the other constant. $\varepsilon_{env}$ represents the fraction of quiescent galaxies within a given environmental overdensity ($\delta$) compared to a reference underdense region ($\delta_0$), while $\varepsilon_{mass}$ signifies the fraction of quiescent galaxies with stellar mass ($M$) relative to the lowest stellar mass at that redshift ($M_0$).

According to \citep{chartab2020}, the environmental quenching efficiency is defined as:
\begin{equation}\label{eenv}
    \varepsilon_{env}(\delta,\delta_0,M,z) = 1 - \frac{f_s(\delta, M, z)}{f_s(\delta_0, M, z)},
\end{equation}
where $f_s(\delta,M,z)$ denotes the fraction of star-forming galaxies with stellar mass $M$ at an environmental overdensity $\delta$, and $\delta_0$ is indicative of an under-dense environment. As defined by \citep{chartab2020}, mass quenching efficiency is defined as:
\begin{equation}\label{emass}
    \varepsilon_{mass}(\delta,M,M_0,z) = 1 - \frac{f_s(\delta, M, z)}{f_s(\delta, M_0, z)},
\end{equation}
where $M_0$ is the threshold stellar mass. Consistent with the methodology outlined by \cite{Kawinwanichakij2017}, we define the lower 25\% of the $\delta$ distribution ($\delta_{25}$) as $\delta_0$. $M_0$ is derived from the stellar mass completeness limit ($M_{min}(z)$).

\begin{figure*}[ht]
    \centering
    \begin{subfigure}
        \centering
        \includegraphics[width=1\columnwidth]{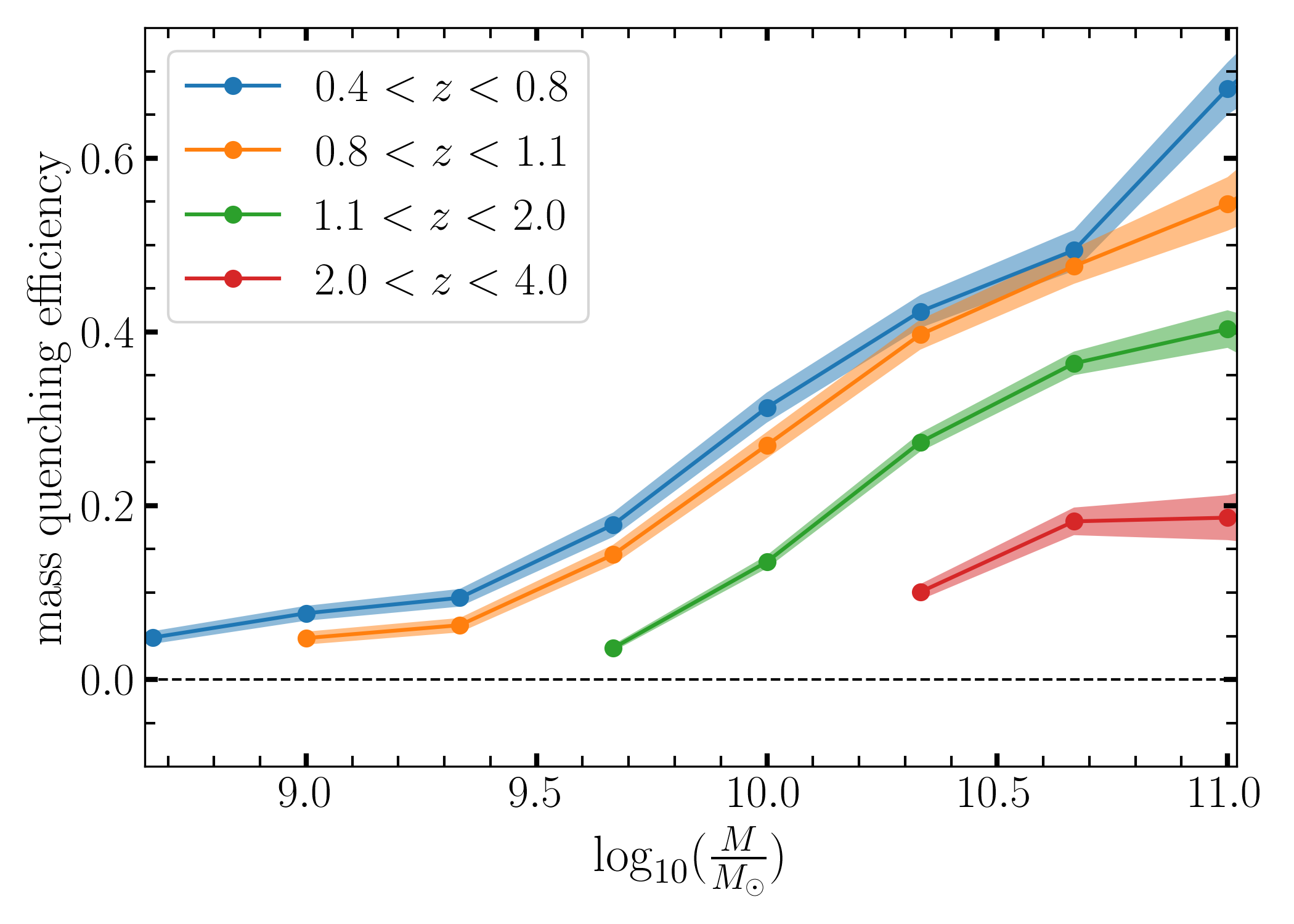}
    \end{subfigure}
    \hfill
    \begin{subfigure}
        \centering
        \includegraphics[width=1\columnwidth]{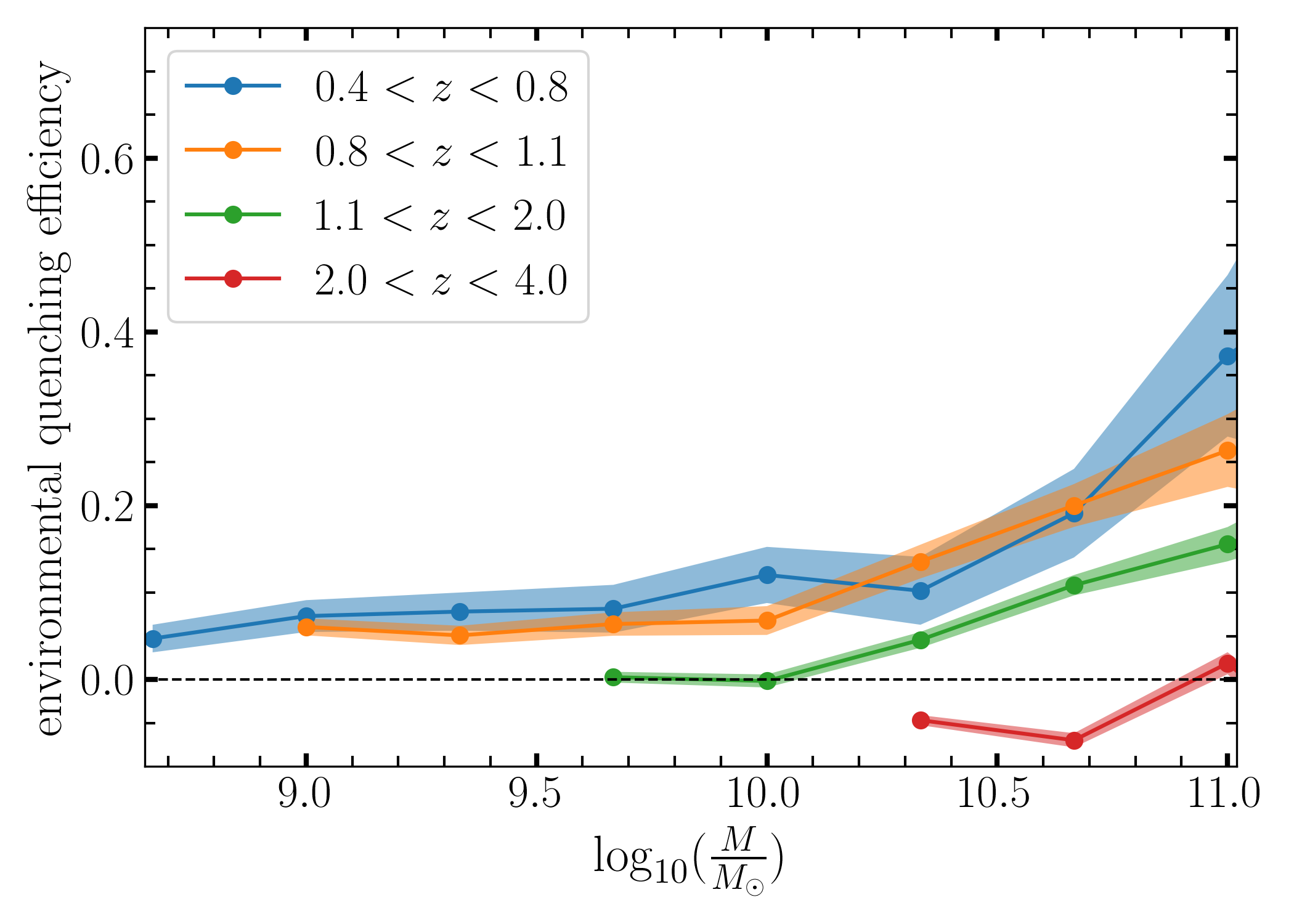}
    \end{subfigure}
    \caption{Quenching efficiencies at different redshifts, illustrating the trends in mass and environmental quenching efficiency across stellar mass bins. The shaded regions represent uncertainties associated with Poisson statistics for the number of quiescent and star-forming galaxies. The Y-axis is limited to -0.1 to 0.75 in both figures.}
    \label{fig:eff}
\end{figure*}

Quenching efficiencies are computed separately for galaxies in overdensities exceeding the 75th percentile of the $\delta$ distribution ($\delta_{75}$) and those inhabiting regions below this threshold. Consequently, we present the stellar mass dependence of mass quenching efficiency, 
$\varepsilon_{mass}(\delta<\delta_{75},M,M_{min}(z),z)$ and environmental quenching efficiency, 
$\varepsilon_{env}(\delta>\delta_{75},\delta<\delta_{25},M,z)$ in Figure \ref{fig:eff}. They are calculated in stellar mass bins of width $\Delta M \sim 0.67 \, dex$. The shaded regions in Figure \ref{fig:eff} depict the uncertainty of the quenching efficiencies, accounting for Poisson statistics associated with the number count of quiescent and star-forming galaxies.

Figure \ref{fig:eff} demonstrate evidence of a monotonic relation between mass quenching efficiency and SM of galaxies with more massive galaxies being more efficient in quenching at all redshifts. The higher quenching efficiency observed in more massive galaxies is likely attributed to their deeper gravitational potential well which leads to shock heating \citep{Birnboim2003,Dekel2006}, and more gas consumption as a result of their higher SFR. In general, internal quenching mechanisms become increasingly effective with higher galaxy mass, with AGN feedback, virial shock heating, and morphological changes playing key roles in suppressing star formation in massive galaxies. These processes are less effective in lower-mass galaxies, where supernova feedback and gas consumption might be dominant factors. The combination of these mechanisms results in a mass-dependent quenching efficiency, leading to the observed trend that more massive galaxies are more likely to be quenched than their lower-mass counterparts. Similar to mass quenching efficiency but to a lesser extent, the efficiency of environmental quenching increases by increasing SM. Figure \ref{fig:eff} shows that mass quenching efficiencies are generally higher at lower redshifts across all mass ranges. However, this trend is less pronounced for environmental quenching efficiency at $z<1$. The negative environmental quenching efficiencies at $2 < z < 4$ indicate that the fraction of star-forming galaxies in dense environments exceeds that in less dense regions. 

In conclusion, as SM increases ($M\gtrsim10^{10}M_\odot$), mass quenching becomes more dominant than environmental quenching mechanisms. This implies that mass quenching is the primary mechanism responsible for ceasing star formation in massive galaxies, while environmental quenching equally contributes to the quenching of lower-mass galaxies, particularly at lower redshifts. 

\section{Comparison with literature} \label{sec:comparison}

In Section \ref{sec:results}, we presented our findings in three redshift epochs. In this section, we investigate scenarios proposed in the literature that can explain these observations and compare our findings with previous studies on the role of mass and environment in galaxy evolution.

%%%%%%%%%%%%%% Literature: Low redshift %%%%%%%%%%%%%%%%%%

Out to $z\sim 1$, the anti-correlation between SFR and environmental density is well-established and reported in numerous studies \citep{Scoville2013, darvish2016, chartab2020, taamoli2024}. Our findings confirm the decrease in average SFR for massive galaxies ($\log (M/M_{\odot}) \gtrsim 10.2$) in all ranges of environments and for lower mass galaxies in high-density environments ($\log(1+\delta) \gtrsim 0.75$). We attribute this observation to the presence of the quiescent population in the overall sample, as these trends diminish in the star-forming sample. \citep{chartab2020} shows that the SFR of massive galaxies ($\log (M/M_{\odot}) \gtrsim 11$) is inversely correlated with the environment at all redshifts. This aligns with our results at $z \lesssim 1$ and supports the scenario in which environmental and mass quenching mechanisms, such as ram pressure stripping, major and minor mergers, and AGN feedback, suppress star formation activity at lower redshifts, in a relatively short amount of time, and form a population of quiescent galaxies \citep{darvish2016}. Based on our results, environmental quenching efficiency increases from 0.1 at $\log (M/M_{\odot}) \sim 10$ to 0.4 at $\log (M/M_{\odot}) \gtrsim 11$, indicating that environmental effects are generally more efficient for massive galaxies, although mass quenching remains dominant. Ram pressure stripping is primarily effective as a quenching mechanism in less massive galaxies (\(\log (M/M_{\odot}) \lesssim 9\)) within the dense cores of clusters \citep{darvish2016}. In contrast, massive galaxies (\(\log (M/M_{\odot}) \gtrsim 11\)) are more likely to host merging companions \citep{Bundy2009}, with major mergers being a dominant factor in the formation of massive, passive galaxies \citep{vanderWel2009}. The higher merger rates expected for massive galaxies in dense environments, due to the increased galaxy number density \citep{Patton2008,Fakhouri2009,sobral2011,Xu2012}, can qualitatively explain the enhanced environmental quenching efficiency observed in these galaxies. Additionally, shock heating from gas accretion in dense environments may further contribute to the quenching of massive galaxies \citep{Dekel2006}. In our star-forming sample, massive galaxies exhibit higher SFR, while their sSFR is lower (Figure \ref{fig:binnedstat}). This can be attributed to the impact of galaxy mass in suppressing the star formation activity of star-forming galaxies. In addition, there is evidence that central and satellite galaxies might be subject to different quenching processes with different time scales, with central going through slower quenching processes in longer time-scales ($\gtrsim 2$ Gyr), whereas satellites may experience rapid quenching in shorter time-scales ($\lesssim 1$ Gyr) after their first infall \citep{Wetzel2013}. Therefore, further studies of the quenching efficiencies in central and satellite galaxies are needed to shed more light on this problem (Taamoli et al. 2025, in preparation).

%%%%%%%%%%%%%% Literature: Mid redshift %%%%%%%%%%%%%%%%%%
At $1<z<2$, several studies report the persistence of $z<1$ trends, anti-correlation between SFR and environmental density \citep{chartab2020}. However, other studies have observed no significant correlation between SFR and environmental density \citep{Scoville2013,darvish2016}, and some have even reported a reversal of the local Universe trends at $z \sim 1$ \citep{elbaz2007, cooper2008}. According to \citep{chartab2020}, the inverse correlation between massive galaxies ($\log (M/M_{\odot}) \gtrsim 11$) and their environment persist out to $z\sim 3.5$ and they do not report a significant environmental dependence for lower mass galaxies at $1.2<z<3.5$. The discrepancy between our results and those reported in \citep{chartab2020} may be due to cosmic variance, as their study was conducted in five smaller, non-contiguous CANDELS fields, whereas our sample covers a larger, contiguous area. Based on our findings, there is no significant correlation between SFR/sSFR and the environment at $1.1<z<1.5$, and there is a statistically significant negative correlation at $1.5<z<2$, but weaker compared to the lower redshifts. At this redshift range, our results are in better agreement with \citep{Scoville2013,darvish2016}. At $1.5<z<2$, SFR is more uniform across the whole mass range compared to the other redshift ranges. However, we report lower sSFR for massive galaxies in both overall and star-forming galaxies at this redshift range. In general, we suggest that the $1<z<2$ epoch can be seen as a transition period between the $z<1$ and $z>2$, where we find the reversal of the SFR environment relation.

%%%%%%%%%%%%%% Literature: High redshift %%%%%%%%%%%%%%%%%%
Beyond $z \sim 2$, trends are even more controversial. While some studies suggest the persistence of the anti-correlation between SFR and density up to $z \sim 3.5$ \citep{chartab2020}, others report a reversal of this anti-correlation at $z\sim 2$ \citep{lemaux_vimos_2022, taamoli2024}. In our analysis, although we note that only a small fraction of galaxies survive the 95\% completeness limit at $3<z<4$, we report a positive correlation coefficient between SFR/sSFR and density at $2<z<4$ for both overall and star-forming samples, even if we take the role of mass into account (see Table \ref{table:corrcoeffs}). Our results support a scenario where galaxies at higher redshifts are more star-forming in rich environments compared to their counterparts in low-density environments. We attribute this to the greater availability of gas through accretion, which initially boosts star formation, along with processes such as tidal effects, higher merger rates, and increased interactions between galaxies that can compress gas and trigger star formation. Additionally, the denser environments may help retain cold gas, further fueling star formation over time. However, at lower redshifts, environmental processes that initially triggered star formation may begin to play a role in suppressing it. Mergers and tidal interactions, which once compressed gas, could eventually lead to gas depletion and quenching. Similarly, ram pressure stripping may become more effective at removing gas in dense environments, potentially reducing star formation activity. This suggests the shifting role of environmental processes at different epochs of cosmic history.

At $z\sim4$, galaxies in dense regions form stars more efficiently than those in low-density regions, whereas in the local universe ($z\sim0$), galaxies in dense regions exhibit lower star formation efficiency compared to those in less dense environments (see their figure 1).

%%%%%%%%%%%%%% Simulations %%%%%%%%%%%%%%%%%%
In addition to observational studies, hydrodynamic galaxy simulations provide insight into the role of mass and environment on star formation activity. \citep{Ghodsi2024} examines this relationship over $0<z<4$ using the cosmological hydrodynamic simulation SIMBA \citep{Dav2019Simba}. Their findings indicate that for low- to intermediate-mass galaxies at $z<1.5$, star formation efficiency is lower in high-density regions compared to low-density regions. However, they observe no significant environmental dependence of star formation efficiency for massive galaxies at any redshift, or for low- to intermediate-mass galaxies at higher redshifts ($z>1.5$). According to \citep{Ghodsi2024}, at $z\sim4$, galaxies in dense regions form stars more efficiently than those in low-density regions, in contrast to trends observed in the local universe (see their Figure 1). These results are consistent with our findings, both indicating that environmental trends observed in the local universe reverse at higher redshifts.

\section{Summary} \label{sec:summary}
In this work, we utilized COSMOS2020 catalog \citep{weaver2022} and density measurements for $\sim 210,000$ galaxies at $0.4<z<4$ from \citep{taamoli2024} to disentangle the contributions of stellar mass and environmental density to the evolution of star formation activity, and to examine the efficiencies of mass quenching and environmental quenching mechanisms at different redshifts. Our findings are summarized as follows:
\begin{enumerate}
    \item At lower redshifts ($0.4 < z< 1$), the mean SFR and sSFR are, on average, smaller for massive galaxies in the overall sample (star-forming and quiescent). For low-mass galaxies, environmental dependence is almost negligible except for high-density environments ($\log(1+ \delta) \gtrsim 0.75$), which is mostly due to the presence of the quiescent population in extreme environments. We report the same environmental trend for massive galaxies too. In the star-forming sample, SFR increases by increasing mass while sSFR decreases. 
    % We report a small, but statistically significant negative correlation coefficient of value $\sim -0.05$ for the overall sample. For the star-forming sample, correlation coefficients are smaller, $\sim -0.02$, with smaller statistical significance for $0.8<z<1.1$.
    \item At intermediate redshifts ($1 < z < 2$), we do not report a strong environmental dependence for the SFR and sSFR. However, we observe a mass dependence in both SFR and sSFR, similar to lower redshift trends: decrease (increase) in average SFR (sSFR) by increasing mass. The Spearman correlation coefficients we reported for this redshift range, are generally smaller and statistically less significant compared to the lower redshifts. This redshift epoch can be considered as a transition phase between low-redshift trends ($z<1$) and the high-redshift trends ($z>2$).
    \item At higher redshifts ($2<z<4$), we report a positive correlation between SFR/sSFR and environment in both overall and star forming samples. At this redshift, the differences between these two samples are smaller compared to lower redshifts. We attribute this positive correlation to the greater availability of cold gas in dense environments, which fuels star formation, as well as increased galaxy interactions and merger rates that can trigger star formation activity.
    \item Mass quenching efficiency increases by increasing stellar mass at all redshifts and environmental quenching efficiency increases by stellar mass out to $z\sim 2$. At $z>2$, the environmental quenching efficiencies are negative, which is due to the larger fraction of quiescent galaxies in over-dense regions, compared to the under-dense regions. For massive galaxies, mass quenching is the dominant mechanism while for lower mass galaxies both environmental and mass quenching mechanisms are effective, particularly at lower redshifts.
\end{enumerate}

This study represents a new milestone in understanding the role of internal and environmental effects on galaxy evolution across cosmic history. These insights are made possible by improved accuracy and precision in physical property estimates and photometric measurements in the latest COSMOS catalog release. Further progress will rely on even deeper future surveys over wider areas to minimize cosmic variance. In the near future, the detection of large samples of structures in wide-area surveys, such as the Euclid Survey and Hawaii Two-0 spectroscopic survey (Euclid Collaboration: \citealt{Mellier2024}, \citealt{zalesky2024}, and \citealt{McPartland2024}) is expected to shed more light on the environmental dependence of galaxy properties. In a forthcoming study (Taamoli et al., in preparation), we are identifying cosmic web components (filaments and clusters) within the COSMOS field to examine the distribution of galaxies across different large-scale structures. This analysis will compare the star formation activity of central and satellite galaxies in galaxy clusters and filamentary structures with their counterparts in voids (field galaxies) to further investigate environmental influences on galaxy evolution.

\section{Acknowledgments}
We are grateful to the anonymous referee for their helpful comments that greatly improved the quality of this work. Some of the data used in this study were obtained at the W.M. Keck Observatory, which operates as a scientific partnership among the California Institute of Technology, the University of California, and the National Aeronautics and Space Administration. The Observatory was made possible by the generous financial support of the W.M. Keck Foundation. The authors wish to acknowledge the profound cultural role of the Maunakea's summit within the indigenous Hawaiian community. This work is based on data products from observations made with ESO Telescopes at the La Silla Paranal Observatory under ESO program ID 179.A-2005 and on data products produced by CALET and the Cambridge Astronomy Survey Unit on behalf of the UltraVISTA consortium. This work is based in part on observations made with the NASA/ESA Hubble Space Telescope, obtained from the Data Archive at the Space Telescope Science Institute, which is operated by the Association of Universities for Research in Astronomy, Inc., under NASA contract NAS 5-26555. B.M. and S.T. acknowledge support from the National Science Foundation under Grant No. 2206813.

\bibliography{Ref}{}
\bibliographystyle{aasjournal}
\end{document}